# Blockchain Security by Design Framework for Trust and Adoption in IoT Environment

Gohar Sargsyan, Nicolas Castellon, Raymond Binnendijk, Peter Cozijnsen

CGI, The Netherlands, Email: gohar.sargsyan@cgi.com, nicolas.castellon@cgi.com, raymond.binnendijk@cgi.com, peter.cozijnsen@cgi.com

*Abstract*—With the recent advances of IoT (Internet of Things) new and more robust security frameworks are needed to detect and mitigate new forms of cyber-attacks, which exploit complex and heterogeneity IoT networks, as well as, the existence of many vulnerabilities in IoT devices. With the rise of block chain technologies service providers pay considerable attention to better understand and adopt blockchain technologies in order to have better secure and trusted systems for own organisations and their customers. The present paper introduces a high level guide for the senior officials and decision makers in the organisations and technology managers for blockchain security framework by design principle for trust and adoption in IoT environments. The paper discusses Cyber-Trust project's blockchain technology development as a representative case study for offered security framework. Security and privacy by design approach is introduced as an important consideration in setting up the framework.

## I. INTRODUCTION

The recent advances in IoT (Internet of Things) requires more robust frameworks and approaches for services and systems to detect and mitigate more complex forms of cyber-attacks exploiting the complexity and heterogeneity of IoT networks. The increase in inter collectivity of consumer IoT devices rises more risks for the users. The rise in cyber threats and cybercrime incidents, given these products come with different levels of embedded security features depending on the country of their origin or sale, the lack of standardisation in the communication protocols and practices followed in the operation of Internet of Things (IoT) devices as well as the absence of monetary incentives for manufacturers to fabricate more secure devices [1]. Designing new services and systems requires more careful consideration of all aspects of security. With the current rise in popularity of blockchain, more organizations aiming at designing solutions and services for consumers in IoT environments are beginning to consider this technology to innovate their own IT environments in the first place and then the services they offer. With every new technology, security risks are amplified or diminished depending on its characteristics. Security and Privacy by design approach became even more important due to new and more strict European Union legislations on data protection and security. This paper offers a blockchain security by design framework for IoT environments. This framework is offered to be used by decision makers in organizations that are planning to adopt blockchain technology. The framework is meant to be a high-level practical guide of the top security concerns an organization should consider when starting their own blockchain application or migrating a current application to this new environment. The paper provides insight on Cyber-Trust project for better understanding the proposed framework. One of the key considerations is security by design approach which sets strong basis from the start of any technology/service development.

## II. BLOCKCHAIN TECHNOLOGY AND SECURITY

Blockchain has become a fast-rising technological trend. Though the origins of its popularity are in cryptocurrencies, we are now starting to appreciate this technology for the changes it can bring in many different technologies, systems, services and organisations' own IT environments. Blockchain technology provides some advantages that are not available in conventional databases, IT systems or applications. It offers the possibility to avoid a central authority, eliminates intermediaries, provides real-time settlement, reduces operational costs, and has high levels of transparency. These are just some of the potential advantages that this new technology can contribute to any new technologies and environments. Although since blockchain can be used to enhance security, the challenge can potentially be the performance and scalability. With every new technology, also come new perspectives to security risks. In this way, blockchain technology is no different than any other modern technology- such as Cloud (Edge) computing or the IoT. All technology is vulnerable to security risks.

Identifying risks for new technologies entails examining the technology and assessing how it can amplify or reduce certain risks. As more organizations begin to consider blockchain technology as a possible solution to offer tot heir customers or innovate their own IT infrastructure, applications and services, it is important to consider the security risks of this new environment. For blockchain, this will concern risks brought by its key characteristics. These characteristics are its distributed nature, its cryptographic seal, its immutability, its transparency, and scalability to name a few. These new characteristics are at the core to understand what the security risks, the implementation and design compexity are for this new technology.

It is important to note that blockchain technology is not a new technology. It's often used to describe distributed ledger technologies (DLT). However, understanding the security risks of the complexity of DLT/blockchain with the recent advancement and rapid technological evolution, is at rather early stage.

For the purposes of this framework, we have defined blockchain technology as an immutable distributed ledger of cryptographically signed sets of records or transactions that a number of parties want to continuously use and extend. These updates, or sets of updates, are saved on the ledger in the form of a "block". Each of the new updates to the ledger is linked to the preceding block and is timestamped, establishing an order for the records. Blockchain technology makes use of two proprietary characteristics - the use of validation rules and their enforcement. Validation rules define the conditions in which the records and blocks will be included in the blockchain and the enforcement of validation rules work in the way of an algorithm or protocol that enforces rules that have been entrusted by all parties that contribute data to the blockchain.

*New Perspectives to Security Concerns*

As previously stated, blockchain technology presents a new perspective on security issues. Two considerable attributes of this technology are it scalable and decentralized natures. These two characteristics of blockchain provide major advantages for its use. These advantages may also present strains on the security of the technology. The Blockchain Trilemma is a concept that exemplifies how the characteristics of this technology may pose a strain on its security. The blockchain trilemma is a concept that explains how there is a tension between the scalability, decentralization, and security of the technology. Though this concept is not widely agreed upon, it will be used as an analogy to explain how the different unique characteristics of this technology have consequences on the security of the application.

**Decentralization**: This is the attribute at the core of blockchain and the main tenant upon which communities around this technology were built. Its decentralized nature means there is no central body that is in control of the information being handled. This means it is censorship-resistance and allows for a nearly democratic participation of users in the ecosystem.

**Scalability**: This refers to the capability of a single node on a blockchain network to handle a growing amount of transactions per second and thus be enlarged to accommodate that growth. A node can be considered scalable if it is capable of increasing the total output under an increase in transactions per second when it is scaled horizontally or vertically. Scalability can be done horizontally, instantiating the same node again so two or more nodes can handle the increased load. Scalability can also be achieved vertically by adding more resources such as additional memory or Computer Processing Units (CPU) to the single node.

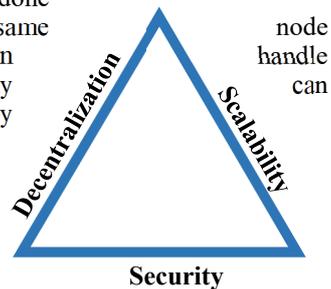

**Security**: This attribute concerns the risks that particular blockchain technology is susceptible to. In a general sense, the security concerns the Confidentiality, Integrity, and Availability of the technology. For blockchain, confidentiality means the authentication of the user or node on the chain; integrity means the data on the chain is immutable and authentic, and availability means the reliable use of the data stored and handled by the blockchain.

The blockchain trilemma suggests that increasing any two of these attributes will have a decrease on the third. Choosing to have a highly scalable blockchain may mean the widening of the attack surface, while decentralization means losing the control and authority over data, on the other handm it means distributed trust. Though these are presented as security risks, these characteristics may make a chain more secure, such as scalability providing more resilience for the application and decentralization spreads the risk of a single point of failure. Taking this dilemma into account, we encourage the user of blockchain technology to use the security of the blockchain as a parameter to measure the attributes and characteristics of this technology, especially when using data linked to personally identifiable information.

*Do you Really Need a Blockchain?*

Blockchain can be simply described to be the orchestration of three technologies- the Internet, private key cryptography and a protocol governing incentives [2]. This all results in a secure system for digital interaction without the need for a trusted third party to facilitate digital relationships. In this way, blockchain technology should be seen as a consortium of current technologies applied in a modern innovative way. As mentioned earlier, this technology is not suitable for all use-cases. In order to determine if blockchain technology is ideal for the IT system or process in question, we suggest using the diagram bellow developed by IEEE [3]. The Figure 1 walks the user through the different considerations to take into account when wanting to adopt blockchain technology more generally. These considerations include the satisfaction with using traditional databases, the number of participants that will contribute data, the level of trust among participants, and the level of privacy and control needed over the data.

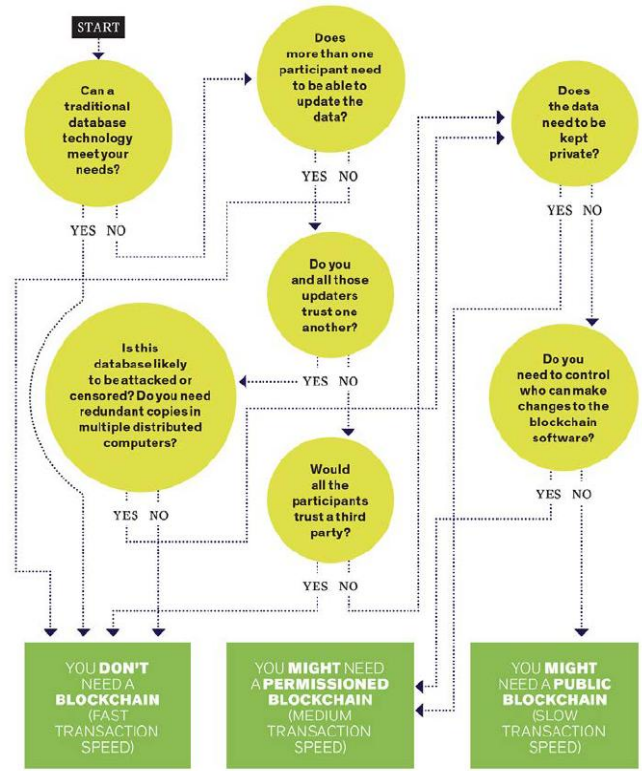

Figure 1: IEEE blockchain decision Tree (2017)

*Consensus Mechanisms*

Blockchain technologies make use of consensus mechanisms to achieve an agreement on a single data value without a centralized authority. Two of the most prominent consensus mechanisms are known as Proof of Work (PoW) and Proof of Stake (PoS). There are many other widely used consensus mechanisms at the moment, including Proof of Identity, Proof of Capacity, Proof of Burn, and Proof of Authority (PoA). Consensus mechanisms ensure that all transactions within a block are agreed upon before adding a new block. As part of this wider verification, blockchain technology makes use of miners that create new blocks and to verify these transactions, very much in the same way we may hire an accountant to review financial information. Miners are selected in accordance with the chosen consensus mechanism, and the miners who successfully respond will verify transactions and also create new blocks on the blockchain. Various consensus mechanisms will do this in different ways.

**Proof of Work** does this by letting miners solve encrypted puzzles. The first miner to solve the encrypted puzzle will verify the transaction, create a new block, and announces the solution to the entire network. In return for this work, the miner gets a reward in the form of an amount of the cryptocurrency being transacted. Without the reward system, miners would not be willing to solve the puzzles, so it is important to be aware of the importance of the reward system. Hardware to mine transactions is expensive and requires a significant amount of electricity to power. This leads to miners operating in consortiums known as Mining Pools. These offer miners the opportunities to pool resources to mine a block, spread the risks, and split the rewards.

**Proof of Stake** differs significantly from a proof of work system. Instead of building blocks through work output, the share or stake in a cryptocurrency determines the creator of a block. In other words, the bigger the share that a miner owns, the more mining capabilities a miner will have. This allows a miner to only mine a percentage of the transactions that are similar to its own share.

*Public, Private and Hybrid Blockchains*

Blockchain technology can be explained in terms of access to a transaction, defining public, private, and hybrid blockchains, and can be defined by access to transaction processing creating the distinction between permission and permissionless blockchainblock chain network.

**Permissionless Blockchain**: In permissionless networks, any user is able to join and begin interacting with the network, such as submitting transactions, adding entries to the ledger, running nodes on the system, and verifying transactions.

**Permissioned Blockchain**: In a permissioned blockchain, the network owner decides who can join the network and only a few members are allowed to verify blocks.

**Public Blockchain**: A public blockchain has entirely an open read access and anyone can join and write in the network. Public blockchains often work with Proof of Work consensus mechanisms to incentivize participation.

**Private Blockchain**: A private blockchain often is the opposite of a public blockchain and only authorized participants have read access and can write and join the network. Often this requires an invitation to join, subsequently either the network starter or a set of rules put in place, determine if someone is fit to join.

**Hybrid Blockchain**: A hybrid blockchain, also known as a consortium blockchain, uses attributes of both private and public chains. It refers to a closed environment in which various parties work together in sharing data and transactions. Members can also determine which transactions can remain public and which have to be restricted to a smaller group of members. The Table 1 below provides a quick overview of the differences between public and private blockchain types and their characteristics.

|  | Public | Private |
|---|---|---|
| Examples | Bitcoin and Ethereum | Hyperledger-Fabric and R3 Corda |
| Consensus algorithm | Commonly used are Proof of Stake and Proof of Work | Agreed upon with pre-defined rules. Proof of Authority mostly used in the Netherlands |
| Scalability of the network (Txs/second) | Low | High |
| Participation in network | Mostly Permissionless. Users are free to join | Mostly Permissioned. A defined group of participants |
| Development | Determined by the community | Controlled by a central party |
| Privacy when using personal data | Not recommended. | Not recommended. Links to data through blockchain is safe up to a certain degree. |
| Identity of the nodes in the network | Anonymous or Pseudonymous | Known identities |

Table 1: Comparison between Public and Private blockchain

*Blockchain Security Framework*

Like all other technologies, blockchain faces a number of security risks that are amplified and minimized accordingly to its unique characteristics. An example of this can be seen in blockchain's consensus mechanism, where it both amplifies and reduces different security risks. In terms of amplifying threats, consensus mechanisms may make certain types of blockchains vulnerable to a 51% attack where an attacker can overpower the network and effectively monopolize and control the application. By controlling the network, attackers would be able to prevent new transactions from gaining confirmations, allowing them to halt payments between some or all users. They would also be able to reverse transactions that were completed while they were in control of the network, meaning they could double-spend cryptocurrencies. In terms of minimizing security risks, this attribute ensured that altering data on a chain is significantly more difficult as the data has been encrypted and cross-checked by other peers in the network. There are several more examples of this sort, where blockchains specific characteristics can reduce and at the same time increase security risks. For this reason, it is recommended to assess this technology with a minimum set of security controls.

This minimum set of controls take care of common security risks ranging from operational such as access control and secure system development, to strategic such as security policies for your organization. If an organization is not developing or maintaining information systems, it is recommended to have a basic level of understanding of what are common security good practices. This understanding allows organizations to challenge IT suppliers on how they implement security controls for the applications that are requested. For a high-level overview of these security controls, it is recommended to use ISO/IEC 27001:2013 or NIST version 1.1 as a baseline. This framework presents 14 security considerations for secure blockchain applications, of which are divided into four categories:

**Blockchain specific**:

This category will describe security issues that are most amplified by blockchain technology. The issues presented in this category are not unique to blockchain technology but are amplified by the technologies characteristics. These will include smart contracts, forks, cryptographic algorithms, and cryptographic key management.

**Network and Infrastructure**:

This category will describe how blockchain should be considered for operations and the general IT infrastructure. These considerations will consist of access control, scalability, intrusion detection, targeted attack resistance, and data propagation attack resistance.

**Operational and Organizational:**

This category will highlight security issues that affect an organization at an operational and organizational level. These security considerations are not unique to blockchain, but must not be forgotten when implementing or adopting this technology. These include operations and communications security, system acquisition, development and maintenance, asset management, human resource security, and supplier relationships.

**Management-level:**

This category will highlight considerations for an organization's management level. They are also not unique to this technology but are crucial for establishing a culture of secure development, implementation, and operation of this technology. These include organization of information security, information security policies, and external and internal compliance.

The Table 2 gives an overview of the level of influence an organisation can have in mitigation the blockchain security considerations[4].

|  |  | Public | Private |
|---|---|---|---|
| Blockchain Specific | | | |
| 1 | Security of Smart Contracts | + | + |
| 2 | Forks | - | + |
| 3 | Crypto Algorithms | - | + |
| 4 | Crypto key management | + | + |
| Network and Infrastructure | | | |
| 5 | Access control | - | + |
| 6 | Scalability | - | + |
| 7 | Intrusion Detection | - | + |
| 8 | Targeted attack resistance | - | + |
| 9 | Data Propagation attack resistance | - | + |
| Operational and Organizational | | | |
| 10 | Operations & Communications Security | - | + |
| 11 | System Acquisition, Dev and Maintenance | + | + |
| 12 | Asset Management | + | + |
| 13 | Human Resource security | + | + |
| 14 | Supplier Relationships | - | + |
| 15 | Incident Management | - | + |
| Management Level | | | |
| 16 | Organization of InfoSec | - | + |
| 17 | Information Security Policies | + | + |
| 18 | External/Internal Compliance | + | + |

Table 2: Comparison between Public and Private blockchain

*Blockchain technology Risks when Migrating*

Migrating an application or process to a blockchain architecture will require an additional list of topics to be considered. Though blockchain has attributes that make it different than other technologies and architectures, it should be assessed like any other technology. The following is a list of comprehensive operational security risk considerations. It is important to note that the considerations have been formulated under the assumption that organizations will be adopting a blockchain technology and not developing a proprietary chain: choosing the right blockchain, special considerations for testing, awareness and training, contingency planning, simplicity as a security measure and privacy. [4].

*Considerations for Privacy*

It is a current trend for privacy concerns in Europe to be automatically linked to the General Data Protection Regulation (GDPR), which became directly applicable in all member states on 25 May 2018. Given its importance, we will focus on illustrating the applicability of the GDPR, understanding the roles of Data Processor and Data Controllers in this context, and the risks to personal data [5]. This chapter will take a closer look at the roles of the data processor and data controller, the preferred type of blockchain in terms of privacy, the rights of the data subjects in the context of a blockchain application, and will discuss hashes in the context of the GDPR.

The GDPR poses serious challenges for organizations that have to comply in order to avoid fines. Blockchain technology is not exempted from this obligation if personal or pseudonymous data is involved in the process. One has to be aware of the fact that the GDPR still causes uncertainty about the interpretation of certain articles in it. Organizations face the same challenges with blockchain applications. When considering blockchain technologies, it is important to consider the relationship between controller and processor and the user's rights.

**Controller vs. Processor**

The first main concern lies in defining the roles of controller and processor for the blockchain application. In the GDPR the controller is defined as the natural or legal person, public authority, agency or other body which, alone or jointly with others, determines the purposes and means of the processing of personal data [5]. The processor can be defined as the natural or legal person, public authority, agency or other body which processes personal data on behalf of the controller [5].The processing of personal data within a blockchain presumes that there is no hierarchical relationship between the participants. Each participant is therefore equal and able to contribute and make use of the date as seen fit [6]. If there are other agreements in place, this could prove to be the exception.

For blockchain applications, a controller can be defined as the participants of a blockchain who have the right to write on the chain and who decide to send data for validation by the miners [7]. More specifically, a controller can be more closely defined as a participant that is seen as a natural person that processes personal data related to a professional or commercial activity or when a participant is a legal person that registers personal data in a blockchain [7]. In other words, the participants that define the purpose and means of processing are the controllers, thus excluding miners from being a controller. The controller has different obligations under the GDPR, such as reporting a data leak. If a group of participants decides to carry out processing operations with a common purpose, this would lead to practical issues with regard to governing these responsibilities. This should be addressed in various ways. One way to do this is by identifying one participant as the decision maker by reaching an agreement on how to govern as joint controllers. Another way to achieve this is by creating a legal persona such as an economic interest group or association [7]. This issue can likely be solved within a blockchain that is governed by one or a few parties.

If parties that do not necessarily exchange personal data, but are contributing as nodes to the blockchain network, it can be assumed that these parties can be considered processors [7]. In other words, one could say that all the nodes that are not specifically defined as being controllers could be considered processors since they all contribute as a node to the processing, creation, and maintenance of the data on the chain. Consequently, all the controllers have to enter into a processing contract with the processors. In a small private blockchain this is quite manageable, yet in a larger private or public blockchains, this is a more complicated matter. Organizations should be aware that there is no legal precedence on this matter, thus European case law could lead to different interpretations. For this same reason, it is currently unclear what the definition of processors could mean for public blockchains and what legal obligations controllers have with regard to processors.

From a privacy perspective, permissioned and private blockchain applications are the safe choice for organizations wanting to adopt this technology. These two types of blockchain make it easier to identify controllers and processors. In return, this makes the governance of legal obligations for controllers and processors more manageable, as well as taking care of the contractual obligations between controllers and processors. It is very difficult to identify all the controllers and processors in a public blockchain, making it questionable if it is legally possible to adhere to the GDPR when using a public blockchain.

### Data Subject Rights

An important component of the GDPR concerns the data subject rights. A data subject has six different rights under the GDPR: the right of access and rectification, the right of erasure, the right to restrict processing, the right of data portability, the right to object, and the right to not be subject to automated processing [5].

We will be focusing on three of these rights and how they present challenges to the use of blockchain technology. These three rights are the right to erasure, the right to rectification, and the right to limit processing. Data subject rights are at the core of the GDPR and present the biggest concerns as there are no exemptions to their compliance.

### Regarding Hashes and Personal Data

It is worth mentioning that at this moment, a hash is considered to be personal data. The Dutch DPA provides three important reasons why a hash is considered as personal data [8]. Firstly, because the source data is often still available and the hash is then used in combination with a linking table; this leads to pseudonymization and not anonymization of the data. Secondly, it is theoretically feasible that hash values can be reproduced using a brute-force attack. Although it is rather difficult to brute-force a hash value back to the original data, this notion postulates that is technically possible. Thirdly, organizations often store the hashes with other additional information. The combination of those two could make it possible to link a person to a hash. Two of the three mentioned factors can be limited by fully separating the hash from the source information and other additional information, which is a measure that is mentioned before when discussing the data subject rights.

### Compliance Beyond the GDPR

All in all, this section highlights some specific issues to be taken into account when discussing personal data processed on a blockchain application. Besides GDPR compliance, there are also other legal considerations that should be taken into consideration when working with a blockchain application in general. A whitepaper from Pels Rijcken & Droogleever highlights a few of these legal considerations, such as how to define the applicable national law for an international blockchain, how to define legally the ownership of a blockchain, legal issues with regard to identity within a blockchain – which especially applies to public blockchains, legal issues with regard to smart contracts, and legal issues concerning the monitoring of blockchains.i It is sensible to delve into this matters, to make sure that a blockchain adheres to certain legal obligations. In addition, it is crucial to always take specific legislation into account, which is already applicable to the sector in which the blockchain will be used [6]. Taking these considerations into account will bring organizations one step closer to adhere to its legal obligations when using a blockchain application.

### III. BLOCKCHAIN AND IOT

Quite some time blockchain is increasingly being looked upon in an IoT context. Blockchains have many benefits. They are not the answer to all challenges of the digital economy as sometimes is said but it's certain that they will play an increasing role in the IoT. The intersection of blockchain and IoT is becoming more emerging topic. What happens when the distributed IoT meets Distributed Ledger Technology. As IoT applications are by definition distributed it's only normal that the distributed ledger technology, which blockchain is, will play a role in how devices will communicate directly between eachother (keeping a ledger and thus trail of not just devices but also how they interact and, potentially, in which state they are and how they are 'handled' in the case of tagged goods). Blockchain is designed as a basis for applications that involve transaction and interactions. These can include smart contracts, which are contracts that are automatically carried out when a specific condition is met, for instance regarding the conditions of goods or environmental conditions or other smart applications that support specific IoT processes. This way blockchain technology can improve not just compliance in the IoT but also IoT features and cost-efficiency.

### IV. CASE STUDY: CYBER-TRUST PROJECT

The Cyber-Trust project [9] aims to develop an innovative cyber-threat intelligence gathering, detection, and mitigation platform to tackle the grand challenges towards securing the ecosystem of IoT devices. The security problems arising from the flawed design of legacy hardware and embedded devices allows cyber-criminals to easily compromise them and launch large-scale attacks toward critical cyber-infrastructures. Cyber-Trust is designed especially for application in IoT domain. In particular, in order to address the challenges in the forensic evidence collection, preservation and investigation process, for IoT environments in the smart home domain, by exploiting the advanced intrusion detection, distributed ledger technology (DLT) and attack mitigation solutions that are being developed in the context of the Cyber-Trust project.

Cyber-Trust offers sound solution for the monitoring and safeguarding of end-user devices, combining a range of technologies with a loosely-coupled, multi-tier and modular architecture. Limitations and restrictions for its operation are subject to the end-users' setup and compatibility between monitored devices. The method of architecture design applied within Cyber-Trust is the proven Risk- and Cost-Driven Architecture (RCDA) approach. The advantage of applying this method is that it supports just-in-time just-enough architectural and design decision making throughout the whole design process. Concerns and decisions are weighed throughout the design process and stakeholders' requirements are constantly validated against the design. The design process is iterative to ensure high quality results. As a part of a complex solution blockchain security by design principles have been applied to secure more robust architecture. In addition, legal experts are envolved continuously at all stages: requirements, design and implementation.The project Cyber-Trust is still ongoing and development continues. The consortium partners has already shown remarkable progress in achieving its goals.

The integrated Cyber-Trust prototype solution will be released in early 2020 to undergo two piloting cycles, capturing the needs of its major stakeholders. The first piloting cycle is planned to last four months and will focus on the platform's capability in the detection and mitigation of cyber-attacks and the solution's impact when deployed in a large scale by an Internet Service Provider (ISP). After the end of this phase, the integrated platform will be refined with further fixes and updates, as these emerge from the needs of the ISP. The final version of the platform will then be tested to its full capabilities, for a two-month period, by also involving the second key Cyber-Trust stakeholder, a LEA. Cyber-Trust committed to the delivery of advanced cyber intelligence tools to the open-source community will publish its final product publicly on GitHub.

We will witness the application side of the offered block chain security framework during the execution of the pilots. To date, the partners are confident that this is the way to go and secure more robust architecture of the system.

## V. Conclusion

In principle, with more connectedness of IoT (Internet of Things) new more robust security frameworks are needed to detect and mitigate new forms of cyber-attacks. Among other secure technologies, the importance of blockchain technologies is gaining momentum for technology companies, service providers for their own IT systems and also for applications and services they offer to their customers. In the current papers we introduced the blockchain technology security framework, the security by design considerations in blockchainblock chain technology, trust and adoption principles, security and privacy by design principle [10] and data privacy. The paper is aims to be a high level guide for the senior security officials and decision makers in the organisations, technology, application and service providers. The paper also discusses a real case study, Cyber-Trust project applying and adopting the approach.

## Acknowledgment

This project has received funding from the European Union's Horizon 2020 research and innovation programme under grant agreement no. 786698. This work reflects authors' view and Agency is not responsible for any use that may be made of the information it contains.